\documentstyle[twocolumn,epsf,aps]{revtex}
\begin{document}
\title{Quantum key distribution with 2-bit quantum codes
}
\author{Xiang-Bin Wang\thanks{email: wang$@$qci.jst.go.jp} 
\\
        Imai Quantum Computation and Information project, ERATO, Japan Sci. and Tech. Corp.\\
Daini Hongo White Bldg. 201, 5-28-3, Hongo, Bunkyo, Tokyo 113-0033, Japan}

\maketitle 
\begin{abstract}
We propose a prepare-and-measure scheme for quantum key distribution 
with 2-bit quantum codes.
 The protocol is unconditionally secure under whatever
type of   intercept-and-resend attack.
Given the symmetric and independent errors to the transmitted qubits,
our scheme  can tolerate a bit error rate up to
26\% in  4-state protocol and  30\% in  
6-state protocol, respectively. These values are higher than 
all currently known threshold values for the prepare-and-measure protocols. 
Moreover, we give a practically implementable linear optics realization
for our scheme.
\end{abstract}
{\it Introduction.} Quantum key distribution (QKD) is different from classical cryptography in 
that 
an unknown quantum state is  in principle not known  unless it is 
disturbed, rather 
than the conjectured difficulty of computing certain functions.
The first published protocol, proposed in 1984~\cite{BB}, 
is called BB84 (C. H. Bennett and G. Bras\-sard.) For a 
history of the subject, one may see e.g. \cite{gisin}.
Since then, studies on QKD are extensive.
Strict mathematical proofs for the unconditional security have been given 
already\cite{qkd,mayersqkd,others}.
It is greatly simplified if one connects this with
the quantum entanglement purification protocol 
(EPP)\cite{qkd,shorpre,BDSW,deutsch,squeezed,CSS}.
Very recently, motivated for  higher bit error rate tolerance
and  higher  efficiency, Gottesman and Lo\cite{gl} 
studied the classicalization of EPP with two way communications (2-EPP). 
Their protocol has increased the 
tolerable bit error rate of channel to 18.9$\%$ and 26.4$\%$ 
for 4-state QKD and 
6-state QKD, respectively. Very recently, these values have been upgraded
to 20$\%$ and 27.4$\%$ by Chau\cite{chau}.

\begin{figure}
\epsffile{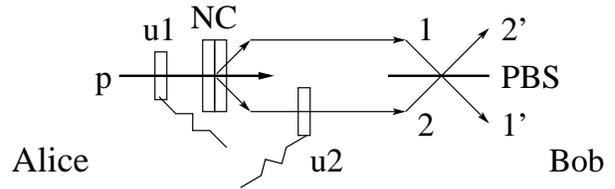}
\vskip 15pt
\caption{ QKD scheme with 2-bit quantum codes.
PBS: polarizing beam splitter.
 NC: nonlinear crystals used in SPDC process,
 p: pump light in horizontal polarization, u1: unitary rotator, 
u2: phase shifter. u1 takes the value of 0, $\pi/2$, $\pi/4$ to produce 
emission state
$|11\rangle, |00\rangle, |\phi^+\rangle$, respectively. u2 can be either
$I$ or $\sigma_z$}\label{pbs}
\end{figure}

\begin{figure}
\epsffile{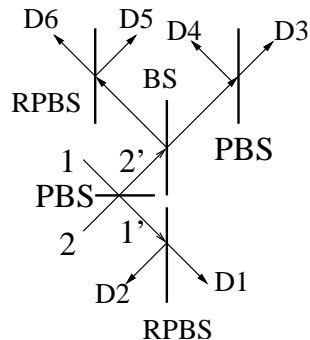}
\vskip 15pt
\caption{Bob's action in QKD scheme of figure(\ref{pbs}). RPBS: Rotated polarizing beam splitter which transmits the state $|+\rangle$ and reflects
state $|-\rangle$). BS: 50:50 beam splitter. D represents for a photon detector.
With RPBS, one may measure the incident beam in 
$\{|+\rangle,|-\rangle\}$ basis.}\label{pbs2}
\end{figure}

The type of prepare-and-measure QKD schemes  is  particularly 
interesting because it does not need the very difficult technique of
quantum storage.
In this paper, we propose a new prepare-and-measure scheme with the 
assistance of 2-bit quantum 
codes. The linear optical realization is shown in Fig.(\ref{pbs},\ref{pbs2}). 
In our scheme, Alice shall send both qubits of the quantum codes 
to Bob, therefore they do not need any quantum storage.  
Bob will first check the parity of the two qubits by the polarizing beam splitter (PBS) and then decode the code with 
post selection. 
The 2-bit code is produced by the SPDC process\cite{para}, see in 
Fig.(\ref{pbs}). 
\\We shall use the representation of
$|0\rangle=\left(\begin{array}{c}1\\0\end{array}\right); 
|1\rangle=\left(\begin{array}{c}0\\1\end{array}\right)$.
We denote  $\sigma_x=\left(\begin{array}{cc}0&1\\1&0\end{array}\right), 
\sigma_z=\left(\begin{array}{cc}1&0\\0&-1\end{array}\right), 
\sigma_y=\left(\begin{array}{cc}0&-i\\i&0\end{array}\right)$.
These operators 
represent for a bit flip error only, a phase flip error only and both error, respectively.
The detected bit (or phase) flip error rate is the summation of $\sigma_x$  (or $\sigma_z$)
error rate and $\sigma_y$ error rate. The Z,X,Y basis are defined by the basis of 
$\{|0\rangle,|1\rangle\},\{|0\rangle\pm |1\rangle\},\{|0\rangle\pm i|1\rangle\}$, respectively.
\\{\it Main idea.}
 We propose a revised 2-EPP scheme which is unconditionally 
secure and which can further increase the 
thresholds  of error rates given the independent channel errors.
We propose to let Alice send Bob the quantum 
states randomly chosen from $\{\frac{1}{\sqrt 2}(|00\rangle + |11\rangle),
\frac{1}{\sqrt 2}(|00\rangle - |11\rangle), 
|00\rangle, |11\rangle\}$. As we shall see, these states are just the quantum
phase-flip
error-rejection (QPFER) code for the BB84 state $\{|0\rangle, |1\rangle,
\frac{1}{\sqrt 2}(|0\rangle + |1\rangle),\frac{1}{\sqrt 2}(|0\rangle - |1\rangle)\}$. 
  
In our 4-state protocol,
the tolerable channel bit-flip and phase-flip rate is raised to $26\%$ for the symmetric
channel with independent noise. (A symmetric channel is defined as the one
with equal distribution of errors of $\sigma_x,\sigma_z,\sigma_y$.)  
Note that the theoretical upper bound of $25\%$\cite{gl} only holds for those
4-state schemes where
Alice and Bob only test the error rate $before$ any 
error removing steps.
However, this is not true with the $delay$ of error test.
Considering the standard purification protocol\cite{BDSW} with symmetric channel,
one may distill the maximally
entangled states out of the raw pairs whose initial bit-flip error and 
phase-flip error are $33.3\%$ .
In our 4-state protocol, we delay the error test
by one step of purification with 2-bit QPFER code. 
This raises the tolerable channel flipping
rates.
\\{\it The QPFER code.}
We shall use the following QPFER code:
\begin{eqnarray}\nonumber
|0\rangle|0\rangle\longrightarrow (|00\rangle + |11\rangle)/\sqrt 2\\
|1\rangle|0\rangle\longrightarrow (|00\rangle - |11\rangle)/\sqrt 2.
\label{pfer}\end{eqnarray}
Here the second  qubit in the left side of the arrow is
the ancilla for the encoding.   
This code is not assumed to reduce the errors in all cases.
But in the case that the channel noise is uncorrelated
or nearly uncorrelated, it works effectively.
Consider an arbitrary state $\alpha |0\rangle_1+\beta|1\rangle_1$ (qubit 1) and
an ancilla state $|0\rangle_2$ (qubit 2). Taking unitary transformation
of Eq(1) we obtain the following 
un-normalized state:
\begin{eqnarray}
\alpha(|0\rangle_{1}|0\rangle_{2}+|1\rangle_{1}|1\rangle_{2})
+\beta(|0\rangle_{1}|0\rangle_{2}-|1\rangle_{1}|1\rangle_{2})
\label{epair}
\end{eqnarray} 
This can be regarded as the encoded state for $\alpha |0\rangle_1+\beta|1\rangle_1$.
Alice then sends both qubits to Bob. In receiving them, Bob first takes a parity check, i.e. he compares
the bit values of the two qubits in Z basis.
Note that this {\it collective} measurement does not destroy the code state itself. Specifically, 
the parity check operation can be done by the PBS in
Fig.(\ref{pbs2}): there, states $|0\rangle,|1\rangle$ are for horizontal and vertical polarization photon states, 
respectively. Since a PBS transmits $|0\rangle$ and reflects $|1\rangle$, 
if 
incident beams (beam 1 and 2) of the PBS are both horizontally polarized or
vertically polarized, there must be one photon on each output beams (beam 1' and 2'); if the polarizations of  two
incident beams are one horizontal and one vertical, 
one of the output beams must be empty. 
After the parity check, if  bit values are different, Bob discards the whole 2-qubit code, 
if they are same, Bob decodes the code. In decoding,  he 
measures qubit 1 in $X$ basis, if he obtains $|+\rangle$, 
he takes a Hadamard transformation $H=\frac{1}{\sqrt 2}\left(\begin{array}{cc}
1 & 1\\1 & -1\end{array}\right)$ to qubit 2;
if he obtains $|-\rangle$ for qubit 1, he takes the Hadamard transformation to qubit 2 and then flips qubit 2 in Z basis.  
Suppose the original channel error rates of $\sigma_x,\sigma_y,\sigma_z$ types
are $p_{x0},p_{y0},p_{z0}$, respectively. Let $p_{I0}=1-p_{x0}-p_{y0}-p_{z0}$.
One may easily verify 
the probability distribution and error type 
for the survived and decoded states (qubit 2) in following table
\begin{center}
\begin{tabular}{rrrr}\hline
JCE  \vline & probability \vline&  decoded state \vline&  
 error type  
\\ 
\hline 
$I\otimes I$  \vline& $p_{I0}^2$  \vline& $\alpha|0\rangle+\beta|1\rangle$   \vline& $I$\\
\hline
$\{I\otimes \sigma_z\}$  \vline& $2p_{I0}p_{z0}$  \vline& $\alpha|1\rangle+\beta|0\rangle$  \vline& $\sigma_x$
\\
\hline 
$\sigma_z\otimes \sigma_z$  \vline& $p_{z0}^2$  \vline& $\alpha|0\rangle+\beta|1\rangle$   \vline& $I$\\
\hline
$\sigma_y\otimes \sigma_y$  \vline& $p_{y0}^2$  \vline& $\alpha|0\rangle-\beta|1\rangle$  \vline& $\sigma_z$
\\
\hline 
$\sigma_x\otimes \sigma_x$  \vline& $p_{x0}^2$  \vline& $\alpha|0\rangle-\beta|1\rangle$  \vline& $\sigma_z$
\\
\hline
$\{\sigma_x\otimes \sigma_y\}$  \vline& $2p_{x0}p_{y0}$  \vline& $\alpha|1\rangle-\beta|0\rangle$  \vline& $\sigma_y$
\end{tabular}
\end{center} 
The first column lists the various types of joint channel errors(JCE) before decoding.  $\{\alpha\otimes\beta\}$  denotes both 
$\alpha\otimes\beta$ and $\beta\otimes\alpha$. 
According to this table, the error rate distribution for the survived raw pairs 
after decoding is: 
 \begin{equation}
  \left\{ \begin{array}{rcl} p_I & = & \displaystyle\frac{p_{I0}^2 +
   p_{z0}^2}{(p_{I0} + p_{z0})^2 + (p_{x0} + p_{y0})^2} , \\ \\
   p_z & = & \displaystyle\frac{p_{x0}^2 + p_{y0}^2}{(p_{I0} + p_{z0})^2 +
   (p_{x0} + p_{y0})^2} , \\ \\
   p_y & = & \displaystyle\frac{2p_{x0} p_{y0}}{(p_{I0} + p_{z0})^2 + (p_{x0} +
   p_{y0})^2} , \\ \\
   p_x & = & \displaystyle\frac{2p_{I0} p_{z0}}{(p_{I0} + p_{z0})^2 + (p_{x0} +
   p_{y0})^2} .
  \end{array}
  \right. \label{errorrate}
 \end{equation}
With this formula, the phase flip error to the decoded states is  obviously reduced. Note that this formula does not hold for the correlated channel 
errors. Even though the noise of the physical channel is uncorrelated,
in carrying out the QKD task, we should not use this formula to $deduce$ the 
flipping rates of the decoded
qubits based on our knowledge of the physical channel noise, i.e.,
the values of $p_{I0},p_{x0},p_{y0},p_{z0}$. But we can choose to directly
test the error rate of the survived and decoded qubits and to $see$ whether
formula (\ref{errorrate}) indeed holds, based on our prior knowledge of
physical channel noise. 
\\{\it Our protocol with linear optical realization.}
 In the BB84 protocol, there are only four different states. Therefore Alice 
may directly prepare random states from the set of  $\{\frac{1}{\sqrt 2}(|00\rangle + |11\rangle),
\frac{1}{\sqrt 2}(|00\rangle - |11\rangle), 
|00\rangle, |11\rangle\}$ and sends them to Bob. This is equivalent to first preparing the BB84 states
and then encoding them by eq.(\ref{pfer}).
We propose the following  4-state protocol with implementation of linear optics in Fig(\ref{pbs}) and Fig(\ref{pbs2}):  
\\
{\bf 1}
Alice prepares N 2-qubit quantum codes with
$N/4$ of them being prepared in   $|00\rangle$ or $|11\rangle$ with equal probability and 
$3N/4$ of them being prepared in  $\frac{1}{\sqrt 2}(|00\rangle\pm |11\rangle)$  with equal probability. All codes are 
put in randomly order.
She records the the ``preparation basis'' as  X basis for code  $|00\rangle$ or $|11\rangle$ ;
and as  Z basis for code $\frac{1}{\sqrt 2}(|00\rangle\pm |11\rangle)$.
 And she records the bit value
of $0$ for the code $|00\rangle$ or $\frac{1}{\sqrt 2}(|00\rangle + |11\rangle)$; bit value $1$ for the 
code $|11\rangle$ or $\frac{1}{\sqrt 2}(|00\rangle- |11\rangle)$. 
She sends each  2-qubit code to Bob. In Fig.(\ref{pbs}), 
any of the above four states
can be produced from the nonlinear crystal by appropriately setting 
the polarization of the pump light\cite{spdcen}.
{\bf 2}
Bob checks the parity of each 2-qubit code in Z basis. 
He discards the codes whenever the 2 bits have different values 
and he takes the following measurement if they have same values:
 he measures qubit 1 in X basis and qubit 2 in either 
X basis or Z basis  with equal probability.
If Bob has measured qubit 2 in X (or ) Z basis, he records 
the ``measurement basis''
as Z (or ) X basis\cite{note0} and we shall simply call the qubit as Z-bit (or X-bit)
latter on.  
If he obtains $|+\rangle|+\rangle$,  $|+\rangle|0\rangle$, $|-\rangle|-\rangle$ or $|-\rangle|0\rangle$, he records
bit value $0$ for that code; if he obtains  $|-\rangle|+\rangle$, $|-\rangle|1\rangle$, $|+\rangle|-\rangle$ or 
$|+\rangle|1\rangle$, he records
bit value $1$ for that code. In our linear optical realization, Bob's 
detections are done by post selection in Fig.(\ref{pbs2}): 
If beam 1' and beam 2' each contains one photon, beam 1 and beam 2
must have the same bit values. Otherwise, their bit values must be different.
This requires Bob to only accept the events of two fold clicking with one 
clicked detector from \{D1,D2\} and the other clicked detector from \{D3,D4,D5,D6\}. All other types of events must be discarded. 
Moreover, according to the above mentioned corresponding rule, to those 
accepted events, clicking of D3 or D4 means measurement in Z basis to beam 2',
corresponding to ``X basis'' for his record; also,  clicking of D5 or D6
means   measurement in X basis to beam 2',
corresponding to ``Z basis'' for his record. 
The two fold clicking of (D1,D6),(D1,D3),(D2,D5) or (D2,D3) corresponds to
bit value of 0; two fold clicking of (D2,D6),(D2,D4), (D1,D5) or (D1,D3)
corresponds to bit value 1.
{\bf 3} Bob announces which codes have been discarded.
Alice and Bob compare the  ``preparation basis'' and  ``measurement basis''
of each bits decoded from the survived codes by classical communication.
 They discard those bits  
whose ``measurement basis'' disagree with ``preparation basis''.
Bob announces the bit value of all X-bits.
He also randomly chooses the same number of $Z-$bits 
and announces their values.
If too many of them disagree with Alice's record,
they abort the protocol.    
{\bf 4}
Now they regard the tested error rates on Z-bits as the bit-flip rate
and the tested error rate on X-bits as phase flip rate. 
They reduce the bit flip rate in the following way: they randomly group all their unchecked bits with each group
containing 2 bits. They  compare the parity of each group. If the
results are different, they discard both bits. If the results are same, they discard one bit and keep the other.
They repeatedly  do so for a number of rounds until they believe that both bit flip rate and phase flip rate can be
 reduced to less than $5\%$ with the next step being taken.
{\bf 5} They then randomly group the remained bits with each group containing $r$ bits. They use the parity of each group
as the new bits.
{\bf 6} They use 
 the classical CSS code\cite{shorpre} to
distill the final key.
\\
Note that in this protocol, Since formula (\ref{errorrate}) is not 
unconditionally true,  Alice and Bob check the bit errors $after$ decoding the 2-bit quantum codes. If the detected errors are significantly 
larger than the expected values calculated from eq.(\ref{errorrate}), they
will abort the protocol. That is to say, if formula (\ref{errorrate}) really 
works, they continue, if it does not work, they abort it.
After any round of bit flip error rejection in step 4, the error rate will be 
iterated by  equation (1) in ref.\cite{chau}.
After the phase error correction in step 5, the new error rate satisfies the inequality of formula (3) of ref.\cite{chau}
provided that $p_I > 1/2$.
The above steps to remove the bit-flip error and phase-flip error are unconditionally true since Alice and Bob 
have paired the qubits $randomly$. Even though the errors of the 
decoded qubits are arbitrarily correlated, the above steps always work
as theoretically expected.  

Given $p_{x},p_{y},p_{z}$, 
if there exits a finite number $k$, after $k$ rounds of bit-flip error-rejection, we can find a $r$ which satisfy
\begin{eqnarray}\nonumber
r(p_x+p_y)\le 5\%\\
e^{-2r(0.5-p_z-p_y)^2}\le 5\%,  
\end{eqnarray}
one can then obtain the unconditionally secure and faithful final key with 
a classical CSS code\cite{shorpre}.

In the 4-state protocol, we don't detect the $\sigma_y$ error for the states decoded from the survived codes,
 therefore we have to assume $p_y=0$ after the quantum parity check and decoding. But we do not have to
assume   $p_{y0}=0$, actually Alice and Bob never test any error rate before decoding in the protocol.
However, $if$ the channel noise is symmetric and uncorrelated, after the quantum decoding, 
both $\sigma_z$ error ($p_z$) and $\sigma_y$
error ($p_y$) are reduced, i.e., the detectable phase error 
rate
has been  reduced in a rate as it should be i.e., eq.(\ref{errorrate}). We then start from the 
un-symmetric error rate with assumption $p_y=0$ and $ p_x,p_z$ being the
detected bit-flip rate and phase-flip rate, respectively. 
After the calculation, we find that the tolerable error rate of bit flip or phase flip is $26\%$ for the 4-state protocol. 
 Moreover, in the case that the channel error distribution itself is $p_{y0}=0;p_{x0}=p_{z0}$,
the tolerable channel error rate for our protocol  is   $p_{x0}=p_{z0}\le 21.7\%$.
The above protocol is totally equivalent to the one based on entanglement purification 
therefore it is unconditionally secure\cite{wang0}. 
Here we give a simple security proof.   
{\it Security proof.}
Consider two protocols, protocol P0 and protocol P. In protocol P0, Alice directly sends Bob
each individual qubits. In protocol P, Alice first encodes each individual qubits by a certain
error rejection code and then sends each quantum codes to Bob. We denote
encoding operation as $\hat E$ and parity check and decoding operation
by $\hat D$. Bob will first check parity of  each code and decode the survived codes. After decoding, Alice and Bob continue
the protocol. Suppose except for the operations of $\hat D$ and 
$\hat E$ everything
else in protocol P0 and protocol P is identical and  operation $\hat D$ or 
$\hat E$
do not require any information of the original qubit itself, then we have the following theorem:
{\bf If protocol P0 is secure with arbitrary lossy channel, then protocol P is also secure.}
The proof of this theorem is very simple. 
Suppose P is insecure. Then Eve. must be able to
attack the final key by certain operation. Denote Eve's attack 
during the period that all codes are transmitted from Alice to Bob 
as $\hat A$. Eve may obtain significant information to final key
with operation $\hat A$ and other operations ($\hat Q$) after Bob receives the qubits.
If this is true, then in protocol P0, Eve may take the operation
of $\hat D \hat A \hat E$ in the same period and then send the decoded states
to Bob, with all other operations identical to those in protocol P.
 (The time order is 
from right to left.) To Alice and Bob, it looks like that they are 
carrying out protocol P0
with a lossy channel now, because Eve will have to discard some of the 2-bit 
quantum codes after the parity check in decoding.
 All final results from  protocol P0 with
attack  $\hat Q\hat D \hat A \hat E$  
must be identical to protocol P with attack $\hat Q \hat A$, 
  since everything there with the two protocols are now the same.
This completes our proof of the theorem.
\\Our
QKD protocol in previous section is just the modified Chau protocol
\cite{chau} with encoding and decoding added. We can regard our protocol
as P and Chau protocol as P0 in applying our theorem.  Since Gottesman-Lo protocol\cite{gl}
or Chau protocol\cite{chau} are all unconditionally secure with arbitrary lossy channel,  we conclude that our protocol must be
also unconditionally secure.
\\{\it 6-state protocol.}
Our protocol can obviously be extended to the 6-state protocol\cite{6state}. In doing  so, Alice just change the initially random codes by
adding $N/4$ codes from $\{\frac{1}{2}[(|00\rangle+|11\rangle)
\pm i(|00\rangle-|11\rangle)]\}$. This is equivalent to 
$\frac{1}{\sqrt 2}(\{|00\rangle
\mp i |11\rangle)\}$. She regards all this
type codes as Y-bits. In decoding the codes, Bob's
``measurement basis'' is randomly chosen from 3 basis, X,Y and Z.
All decoded  X-bits, Y-bits and the same number of randomly chosen
decoded Z will be used as the check bits. Since 
the Hadamard transform
switches the two eigenstates of $\sigma_y$, 
 after decoding, whenever Bob measures  qubit 2
in Y basis, he needs to flip the measurement outcome so that to obtain 
everything the same as that in the 2-EPP with quantum storages\cite{wang0}. 
In such a way, if the channel
is symmetric, Bob will find $p_y\not= 0$. And he will know $p_x,p_y,p_z$
exactly instead of assuming $p_y=0$. This will increase the tolerable error rate accordingly. In the case symmetric physical channel, our 6-state protocol tolerates the flipping rate up to $30\%$. 
\\{\it  Subtlety of the ``conditional advantage''.} 
un-symmetric effective channel. 
Although the advantage of a higher threshold is conditional, the security
of our protocol is $unconditional$. That is to say, whenever our protocol
produces any final key, Eve's information to that key must be exponentially
close to zero, no matter whether Eve uses coherent attack or individual 
attack.
{\it Our protocol is totally different from
the almost useless protocol which is only secure with uncorrelated channel 
noise.}  
There are two conditions for the error threshold advantage:\\
(1) The noise of physical channel should be the type where eq.(\ref{errorrate}) holds;
(2) Eve. is not detected in the error test, i.e., the result of error test must be
in agreement with the expected result given by eq.(\ref{errorrate}).\\ 
Both conditions here are verifiable by the protocol itself.
The second condition is a condition for $any$ QKD protocol. 
The first condition
is on the $known$ physical channel rather than Eve's channel in QKD.  
In our protocol, Eve's attack must not affect 
the error rates detected on the decoded qubits if she wants to hide her 
presence. 
That is to say, if Eve hides her presence, all results about the 
final key of our protocol can be correctly estimated based on the known
properties of the physical channel, no matter what type of attack she has used.
{\it Given a physical channel with its noise being
un-correlated and symmetric and  
higher than the thresholds of all 
other prepare-and-measure 
protocols but lower than that of our protocol, our protocol is the only one that works.} 
In practice,   one may simply separate 
 the 2 qubits of the quantum 
code substantially to guarantee the un-correlation of the physical
channel noise. This is to say, {\it the error
threshold advantage of our protocol is actually unconditional in practice.}
 \\{\it Loose ends in practice.}
Multi-pair emission in SPDC and dark counting of detectors have not been considered.
 We believe these issues  can be resolved along the similar lines in the case
 of BB84 implemented with a weak coherent light source.
\\{\bf Acknowledgement:} I thank Prof Imai H for support. The extended version of
this work has been submitted to QIT-EQIS workshop (Kyoto)\cite{wang0}. 
I thank the 
anonymous referees of the workshop for their comments and suggestions.
I thank Dr B. S. Shi for discussions on the source design. 


\begin{thebibliography}{99}
\bibitem{note0} Here Bob has  omitted  the Hadamard transformation as 
appeared in the EPP.
\end{thebibliography}
\end{document}